\documentclass[aps, prstab, twocolumn, superscriptaddress]{revtex4-1}

\usepackage{graphicx}
\usepackage{dcolumn}
\usepackage{bm}
\usepackage{mathrsfs,amssymb,graphics,subfigure,threeparttable}
\usepackage{amssymb}
\usepackage{enumerate}
\usepackage{amsmath}
\usepackage{fancyhdr}
\usepackage[squaren]{SIunits}

\begin{document}

\preprint{prstab}

\title{Temporal characterization of ultrashort linearly chirped electron bunches generated from a laser wakefield accelerator}

\author{C. J. Zhang}
\affiliation{Department of Engineering Physics, Tsinghua University, Beijing 100084, China}
\affiliation{Laser Fusion Research Center, China Academy of Engineering Physics, Mianyang 621900, China}
\author{J. F. Hua}%
\affiliation{Department of Engineering Physics, Tsinghua University, Beijing 100084, China}
\author{Y. Wan}
\affiliation{Department of Engineering Physics, Tsinghua University, Beijing 100084, China}
\affiliation{Laser Fusion Research Center, China Academy of Engineering Physics, Mianyang 621900, China}
\author{B. Guo}
\affiliation{Department of Engineering Physics, Tsinghua University, Beijing 100084, China}
\author{C.-H. Pai}
\affiliation{Department of Engineering Physics, Tsinghua University, Beijing 100084, China}
\author{Y. P. Wu}
\affiliation{Department of Engineering Physics, Tsinghua University, Beijing 100084, China}
\author{F. Li}
\affiliation{Department of Engineering Physics, Tsinghua University, Beijing 100084, China}
\affiliation{Laser Fusion Research Center, China Academy of Engineering Physics, Mianyang 621900, China}
\author{H.-H. Chu}
\affiliation{Department of Physics, National Central University, Jhong-Li 621, Taiwan}
\author{Y. Q. Gu}
\affiliation{Laser Fusion Research Center, China Academy of Engineering Physics, Mianyang 621900, China}
\author{W. B. Mori}
\affiliation{University of California Los Angeles, Los Angeles, California 90095, USA}
\author{C. Joshi}
\affiliation{University of California Los Angeles, Los Angeles, California 90095, USA}
\author{J. Wang}
\email[]{jwang@ltl.iams.sinica.edu.tw}
\affiliation{Department of Physics, National Central University, Jhong-Li 621, Taiwan}
\author{W. Lu}
\email[]{weilu@tsinghua.edu.cn}
\affiliation{Department of Engineering Physics, Tsinghua University, Beijing 100084, China}

\date{\today}

\begin{abstract}
A new method for diagnosing the temporal characteristics of ultrashort electron bunches with linear energy chirp generated from a laser wakefield accelerator is described. When the ionization-injected bunch interacts with the back of the drive laser, it is deflected and stretched along the direction of the electric field of the laser. Upon exiting the plasma, if the bunch goes through a narrow slit in front of the dipole magnet that disperses the electrons in the plane of the laser polarization, it can form a series of bunchlets that have different energies but are separated by half a laser wavelength. Since only the electrons that are undeflected by the laser go through the slit, the energy spectrum of the bunch is modulated. By analyzing the modulated energy spectrum, the shots where the bunch has a linear energy chirp can be recognized. Consequently, the energy chirp and beam current profile of those bunches can be reconstructed. This method is demonstrated through particle-in-cell simulations and experiment.
\end{abstract}

\maketitle


\section{\label{sec:intro}Introduction}
It is now accepted that electrons can be accelerated by a laser-plasma wakefield accelerator (LWFA) to high energies in a short distance\cite{Tajima:1979bn,Geddes:2004bi,Mangles:2004dd,Geddes:2004tb,Esarey:2009ks}. The ultrashort duration\cite{Lundh:2011js}, small emittance and significant charge make such LWFA generated electron bunches attractive in many applications. For example, compact light sources based on laser-plasma wakefield accelerators have been proposed\cite{Schlenvoigt:2007bg,Kneip:2010kk,Albert:2013jv}. The typical length of electron bunch accelerated by a LWFA is just a few femtoseconds\cite{Lundh:2011js}. Temporal characterization of such short electron bunches is important but extremely challenging. For instance the ability to precisely tailor the current profile of the electron bunch is critical for optimum beam loading while minimizing the energy spread of the injected electron bunch in plasma-based accelerators\cite{Tzoufras:2008jv}.

Parameters of the injected electron bunch are very sensitive to the laser and plasma parameters, which at present makes them less stable than beams obtained from conventional accelerators. To diagnose the length of a bunch accelerated by a conventional accelerator, deflecting cavities\cite{Akre:2002ch} or the electro-optic technique\cite{Berden:2004dl} are frequently used. Although X-band deflecting cavities have shown the capability to resolve beams on a fs-timescale\cite{Akre:2002ch,Behrens:2014ib}, its application to measuring LWFA generated bunches is not yet demonstrated. This is because the transport of ultrashort, tightly focused electron bunches from a plasma accelerator into a deflecting cavity without changing the properties of the bunch is very difficult\cite{Floettmann:2014dlb,Xu:2016kc}. The temporal resolution of the electro-optic technique typically is demonstrated to be on the order of ten $\femto\second$, making it unsuitable for adequately resolving a LWFA produced electron bunch. Coherent transition radiation (CTR) spectrum can also be used to infer the duration of ultrashort electron bunches\cite{Lundh:2011js,Maxwell:2013ek}. However, the accurate measurement of CTR radiation spectrum in the full range from near infrared to IR is challenging. Other methods, such as THz time-domain interferometry\cite{Debus:2010ji}, measuring the oscillation in the energy spectrum resulting from the interaction of the injected bunch with the back of the laser pulse\cite{Kotaki:2015if}, and  measurement based on Compton scattering\cite{Leemans:1996jv} have also been reported.

In this paper, a new method for mapping the current profile of linearly chirped electron bunches generated from a LWFA is proposed. The basic concept is as follows: Due to the dephasing effect that is characteristic of the accelerating electrons in a LWFA, the accelerating electron bunch will catch up with the laser pulse, and will interact with the back of the laser. As it interacts with the electric field of the laser, it will be deflected and stretched along the laser polarization direction. As the electrons exit the plasma they will have a laser phase dependent deflection in the plane of polarization. As a consequence, only a part of the stretched bunch can go through the narrow slit placed orthogonal to the polarization direction in front of the dipole magnet each time the deflection angle of the beam at the plasma exit is zero. In this way, the bunch is now broken up into a series of bunchlets with separations equal to half the laser wavelength, leading to a modulated energy spectrum. If the energy chirp of the electron bunch is linear, the modulated spectrum will contain several peaks with equal spacing in energy. For these shots, the energy chirp and the current profile can be reconstructed. Actually the technique is useful as long as the energy chirp of the bunch is nearly linear. This is because the laser wavelength is being used here as an accurate ruler of the spacing between the bunchelets.

\section{\label{sec:simulation}Simulations}
We first illustrate the proposed method and its usefulness by a high fidelity numerical experiment using the three dimensional (3D), particle-in-cell code OSIRIS\cite{Fonseca:2002wg}. In the simulation, parameters similar to real experimental conditions are adopted to take into account all the major physical effects involved. We use the ionization injection scheme to create the charge to be accelerated within the wake. The neutral gas is a mixture of helium (95\%, atomic percentage) and nitrogen (5\%). The peak density of helium is $3.56\times10^{18}~\centi\meter^{-3}$, which corresponds to a peak electron density of $9\times10^{18}~\centi\meter^{-3}$ under the assumption that the outer five electrons of nitrogen and all electrons of helium are ionized. The density profile of the neutral gas starts with a 382 $\micro\meter$ linear up ramp, followed by a 764 $\micro\meter$ plateau and a 382 $\micro\meter$ linear down ramp. A 3.3 $\tera\watt$, 40 $\femto\second$ (FWHM), p-polarized laser pulse is focused into the neutral gas with a spot size of 11 $\micro\meter$ ($w_0$) to give a vacuum $a_0$ of 0.9. The focal plane of the laser is placed 30 $\micro\meter$ inside from the starting edge of the plateau. A “moving window” is used and the size of the simulation box is $36\times38\times38~\micro\meter$, with a grid size of $25\times190\times190~\nano\meter$.

\begin{figure}[btp]
\includegraphics[width=0.4\textwidth]{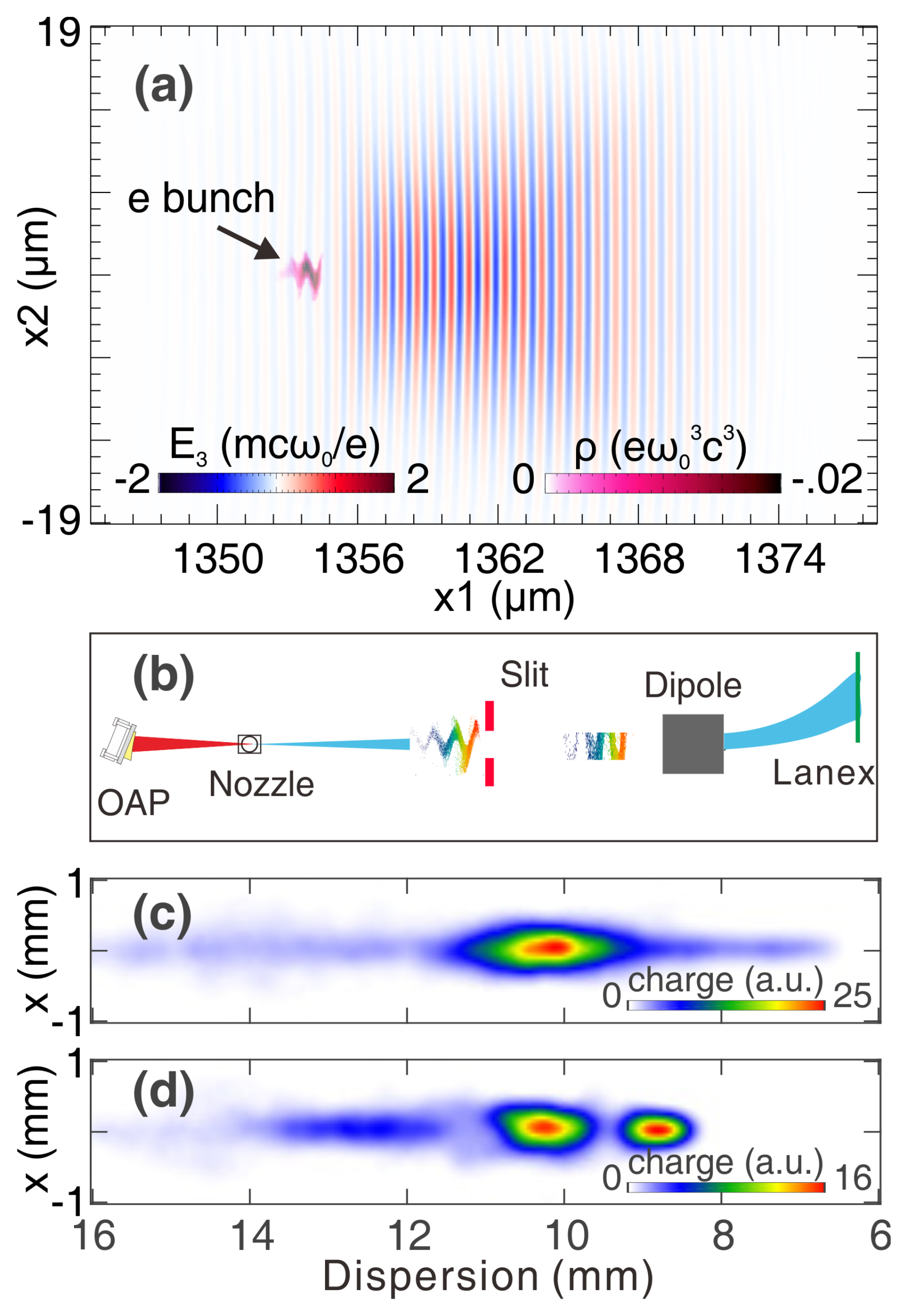}
\caption{\label{fig:1} Output of a synthetic numerical experiment from OSIRIS simulation. (a) Conceptual illustration of the electron bunch interacting with the back of the laser pulse. (b) Schematic of the virtual energy spectrometer with slit which is similar to the experimental set up. The energy spectrums recorded by a virtual screen as seen without (c) and with the slit (d), respectively. The color scales are different in (c) and (d).}
\end{figure}

Figure \ref{fig:1}(a) shows the injected electron bunch (indicated by the arrow) and the laser pulse at the exit of the plasma. The inner shell electrons of nitrogen are ionized near the peak of the laser pulse, quickly slip back in the wake and become trapped. Due to the dephasing effect, the electron bunch catches up with the laser as it propagates and interacts with the back of the laser. Extra momentum modulation along the laser polarization direction is introduced\cite{Shaw:2014kg}, and the bunch will thus be stretched in the transverse direction. The modulation period is equal to the local wavelength of the drive laser, as can be seen clearly in Fig. \ref{fig:1}(a).

Upon exiting the plasma, the electron bunch first passes through a virtual 2-mm-wide slit and then is dispersed in the plane of polarization of the laser by a dipole magnet. The distances between the plasma and the rest of the spectrometer components are similar to those in the experiment to be described later. Since the electron bunch is stretched, only the on-axis portion of the electrons (i.e. those undeflected or very weakly deflected by the laser) go through the slit and make it to the Lanex screen after going through the dipole magnet. In this way, the bunch is now broken up into a series of bunchlets with separations equal to half the laser wavelength, as shown in Fig. \ref{fig:1}(b). Figure \ref{fig:1}(c) and (d) show the energy spectrum recorded by a virtual screen placed behind the dipole, without and with the slit, respectively. As shown in Fig. \ref{fig:1}(c), the energy spectrum is continuous without the slit. And the energy spectrum shows multi-peak features when the slit is inserted, as shown in Fig. \ref{fig:1}(d).

\begin{figure}[btp]
\includegraphics[width=0.4\textwidth]{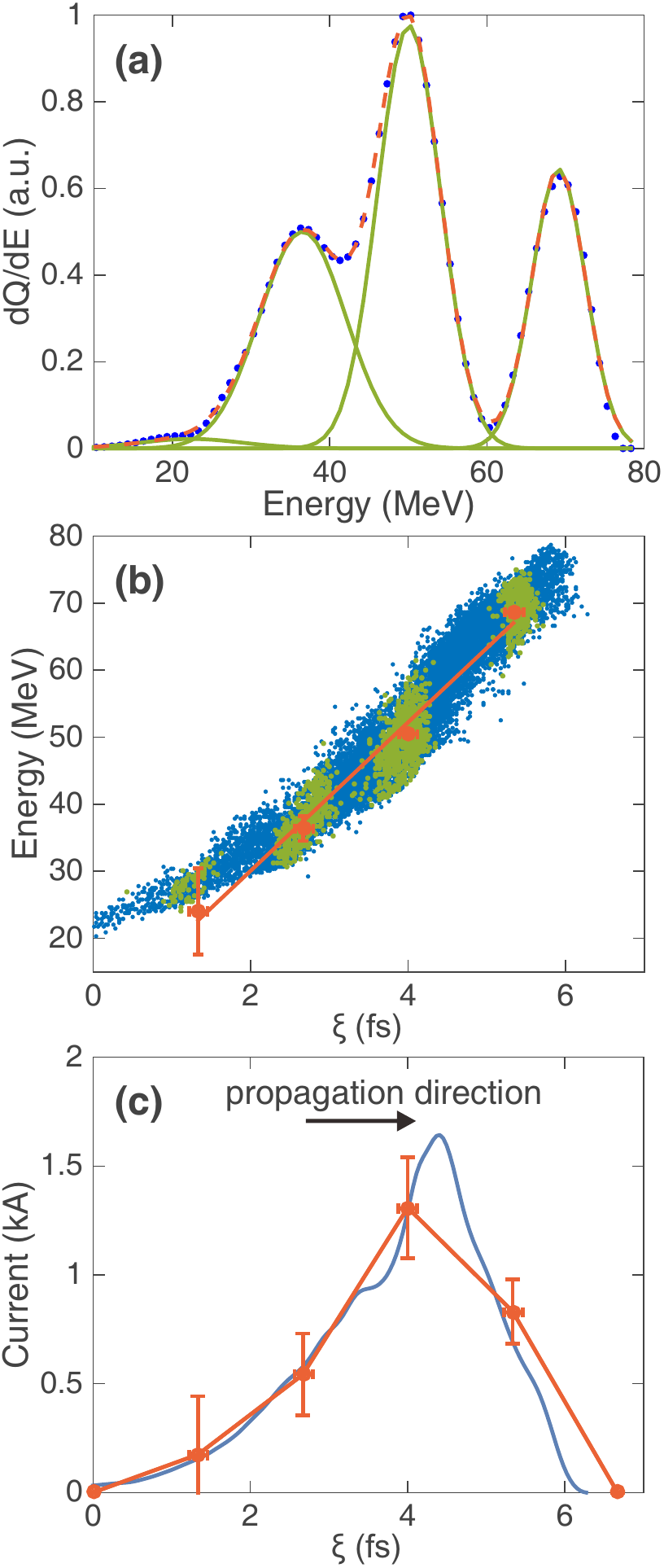}
\caption{\label{fig:2} The blue dots in (a) are the modulated energy spectrum corresponding to Fig. \ref{fig:1}(d). It is fitted by four Gaussian peaks (the green solid lines) and the red dashed line represents the superposition of these four peaks. (b) The longitudinal phase space of the bunch. The blue and green dots represent the electrons received by the virtual energy spectrometer without and with the slit, respectively. The red dots are calculated from (a) and the red line is a linear fit. (c) The beam current profile. The blue line is the current profile directly obtained from the simulation, while the red dots and line is the recovered current profile using the fitting data in (a). The error bars in (b) and (c) arise from the fluctuations induced by the variation of the slit width (from 1.4 mm to 2.2 mm) and the displacement between the bunch center and the slit center (from -0.8 mm to 0.8 mm).}
\end{figure}

Figure \ref{fig:2}(a) shows the deconvoluted energy spectrum (blue dots) corresponding to Fig. \ref{fig:1}(d) after summing the signal in the $x$ direction. The spectrum shows multi-peak feature due to the reason stated above. The nearly equally separated peaks in the energy spectrum indicate that the energy chirp of the bunch is also nearly linear, which will be explained later. The energy spectrum can be fitted by four Gaussian peaks, as represented by the solid green lines. Each peak represents the energy spectrum of the corresponding bunchlet (green dots in Fig. \ref{fig:2}(b)). The central energy and the width of each peak can be easily found from the fitting data. If we assume that the separation between the adjacent bunchlets equals to half the laser wavelength, then the energy chirp can be recovered as shown by the red dots in Fig. \ref{fig:2}(b). The blue and green dots in Fig. \ref{fig:2}(b) represent the electrons received by the virtual energy spectrometer without and with the slit, respectively. As can be seen, the longitudinal phase space data obtained from the simulation (blue dots) is well approximated by the synthetic data (red dots) in Fig. \ref{fig:2}(b). The energy chirp of the bunch can indeed be fitted by a straight (red) line in Fig. \ref{fig:2}(b).

Since the energy chirp is linear, the duration of each bunchlet is proportional to its width in the energy spectrum. As a consequence, the current for each bunchlet is calculated by dividing the charge contained in each peak by the length of the bunchlet. Consequently, the current profile of the bunch is recovered, given that the separation between neighboring bunchlets is half the laser wavelength. The reconstructed current profile is shown by the red dots and line in Fig. \ref{fig:2}(c). The FWHM of the current profile is just 2.2 $\femto\second$ and there is a reasonable agreement between the actual current profile of the bunch (blue curve) and the recovered current profile (red dots and line) from the synthetic energy spectrum.

In the synthetic numerical experiment, the slit width is varied from 1.4 $\milli\meter$ to 2.2 $\milli\meter$ with a stepsize of 0.2 $\milli\meter$. The relative displacement of between the bunch center and the slit center is changed from -0.8 $\milli\meter$ to 0.8 $\milli\meter$ with a stepsize of 0.1 $\milli\meter$. As a result, the red dots in Fig. \ref{fig:2}(b) and (c) show the average of these 85 synthetic shots and the vertical error bars represent the standard deviation. The horizontal error bars arise from the fluctuation of the separation between neighboring bunchlets induced by the displacement between the bunch center and the slit center.

The limit on the slit width is analyzed as follows: To convert the bunch into a series of bunchlets, the slit width should be smaller enough, for example, $\Delta<\sigma_b$ where $\Delta$ is the slit width and $\sigma_b$ is the amplitude of the deflection (the amplitude of the sinusoidal distribution in the laser polarization plane of the deflected bunch as shown in Fig. \ref{fig:1}(b)). The slit width also has a lower limit. Assume that the slice energy spread of the bunch is $\sigma_E$, then the slit should be wider enough so that the energy spread of a bunchlet $\frac{dE}{dz}\frac{\lambda_0}{2}\frac{\Delta}{2\sigma_b}>\sigma_E$, where $\lambda_0$ is the laser wavelength and $dE/dz$ is the energy chirp. Otherwise, the energy spread of the bunchlet will be dominated by the slice energy spread. In this case, the assumption that the width of the Gaussian peak in Fig. \ref{fig:2}(a) is proportional to the length of the corresponding bunchlet no longer holds. Thus the slit width should be $\frac{4\sigma_E}{\lambda_0}(\frac{dE}{dz})^{-1}\sigma_b<\Delta<\sigma_b$. The lower and upper limits for the slit width are 0.4 $\milli\meter$ and 3.2 $\milli\meter$ respectively, calculated with the parameters of the bunch shown in Fig. \ref{fig:2}. In the synthetic numerical experiment, the slit width is varied from 1.4 $\milli\meter$ to 2.2 $\milli\meter$, which covers the middle range between the lower and the upper limits.

The fluctuation of the bunch location affects the separation between neighboring bunchlets. The bunch separation will be half the laser wavelength when the bunch hits the center of the slit like the case shown in Fig. \ref{fig:1}(b). However, if there is a displacement between the bunch center and the slit center, the separation between adjacent bunchlets will be shorter or longer than half the laser wavelength. The density distribution of the deflected bunch in the laser polarization plane can be approximated by $\sigma_b\sin(k_0z)$, where $k_0=2\pi/\lambda_0$ is the wave number of the laser. Thus the change of the bunchlet separation (in unit of $\lambda_0/2$) caused by the displacement between the bunch center and the slit center will be $\frac{2}{\pi}\sin^{-1}(\sigma_x/\sigma_b)$, where $\sigma_x$ is the displacement which is changed from -0.8 $\milli\meter$ to 0.8 $\milli\meter$ in the synthetic numerical experiment. Thus the fluctuation of the bunchlet separation is 0.08 $\lambda_0$ (0.21 $\femto\second$), as shown by the horizontal error bars in Fig. \ref{fig:2}(b) and (c).

As can be seen in Fig. \ref{fig:2}(b), the energy chirp should be large enough to make this technique work. Assume that the deflection due to the laser pulse is significant enough so that the bunch spot size is much larger than the slit width, then the energy chirp should be larger than $4\sigma_E/\lambda_0$, where $\sigma_E$ is the slice energy spread and $\lambda_0$ is the local laser wavelength. Here the slice energy spread is assumed to have a Gaussian distribution. For a bunch with a slice energy spread of 0.5 $\mega\electronvolt$, the energy chirp should be larger than 2.5 $\mega\electronvolt\per\micro\meter$. If the bunch length is 2 $\micro\meter$ and the central energy is 100 $\mega\electronvolt$, then the total energy spread will be 5\%, which is quite close to the typical energy spreads in state-of-the-art LWFA experiments.

\begin{figure}[btp]
\includegraphics[width=0.45\textwidth]{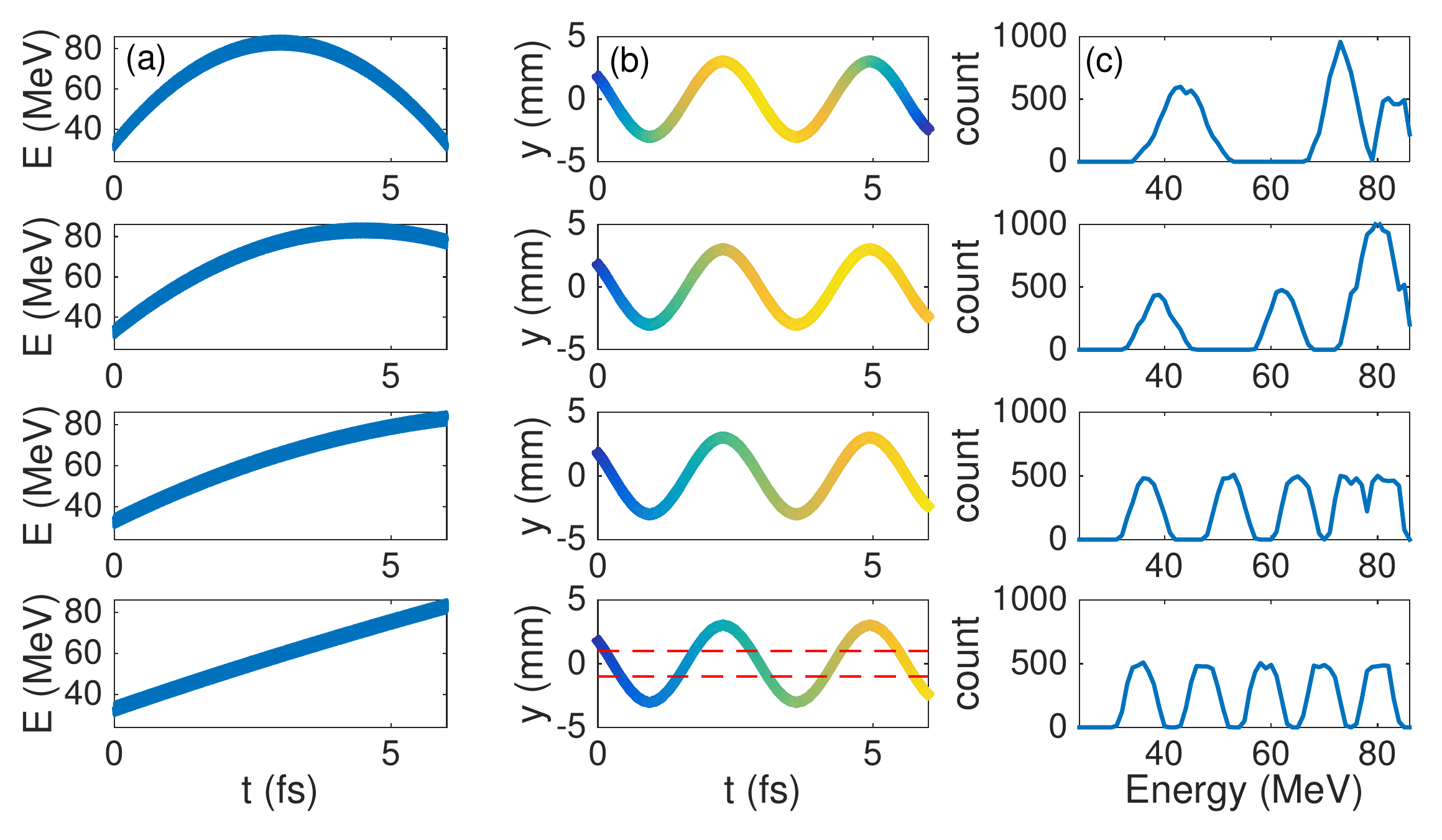}
\caption{\label{fig:3} The left column (a) shows the longitudinal phase space of electron bunches with different energy chirps. The middle column (b) shows the corresponding bunch modulated by the laser, where the bunch length is 6 $\femto\second$ and the laser wavelength is 0.8 $\micro\meter$. The corresponding energy of the electron increases as the color changes from blue to yellow. The amplitude of the modulation is 3 $\milli\meter$, and the width of the slit is 2 $\milli\meter$, as shown by the dashed lines in the last subfigure in column (b). The corresponding energy spectrums measured with the presence of the slit are shown in the right column (c).}
\end{figure}

The linearity of the energy chirp of the bunch can be inferred by analyzing the measured energy spectrum. The equally separated peaks in the energy spectrum indicates that the energy chirp is linear. This is illustrated in Fig. \ref{fig:3} where electron bunches with different chirps are sent through the synthetic spectrometer. The left column of Fig. \ref{fig:3} shows the longitudinal phase space of electron bunches with different energy chirps. The middle column (b) shows the corresponding bunch deflected by a 0.8 $\micro\meter$ laser. The amplitude of the deflection is 3 $\milli\meter$. The red dashed lines in the bottom figure in column (b) shows a 2-mm-wide slit. The corresponding modulated energy spectrums measured with the slit are shown in the right column (c). From these energy spectrums, one can clearly see that as the energy chirp becomes linear, a periodic feature appears. In the bottom figure in column (c), the peaks in the energy spectrum are equally separated, which is an evidence for the bunch to have a linear chirp as shown in the corresponding figure in column (a). In the experiment described below, this criterion, i.e., where the peaks in the energy spectrum are nearly equally separated, is used to select out shots where the bunch is likely to have a linear energy chirp.

\section{\label{sec:experiment}Experiment results}
\begin{figure}[btp]
\includegraphics[width=0.45\textwidth]{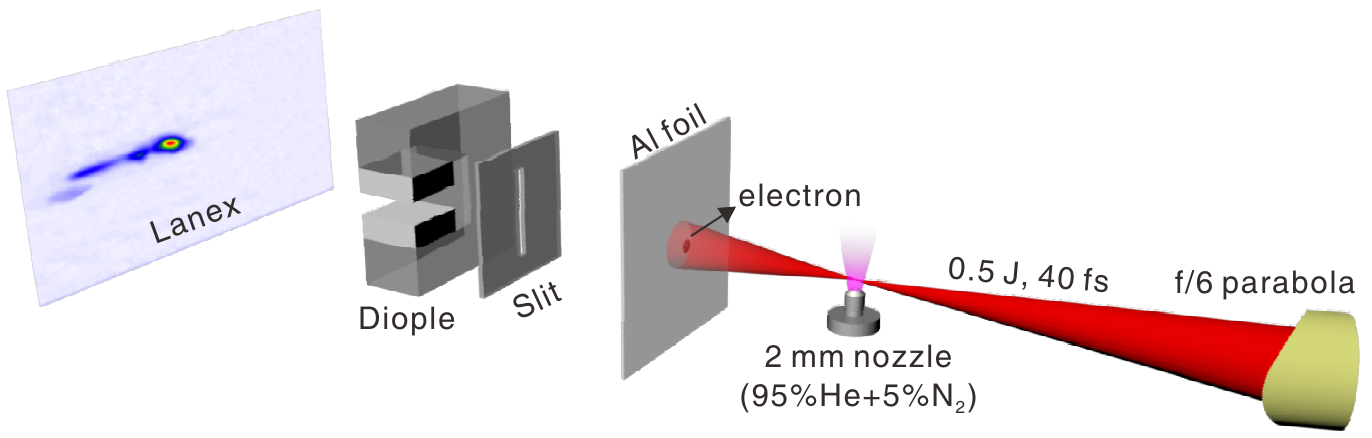}
\caption{\label{fig:4} Experiment setup.}
\end{figure}

The experiment was conducted at National Central University, Taiwan using a Ti:Sapphire laser system with the capability to deliver $\sim30~\femto\second$, up to 100 $\tera\watt$, $\lambda=0.8~\micro\meter$ laser pulses at 10-Hz repetition rate\cite{Hung:2014bm}. The experimental setup is shown in Fig. \ref{fig:4}. The laser pulses are focused by an f/6 parabola onto a gas jet consisting of 95\% helium and 5\% nitrogen, to generate ultrashort electron bunches via ionization injection\cite{Pak:2010eu}. The mean pulse duration of the laser is $40\pm2~\femto\second$ and the laser energy varies from 350 $\milli\joule$ to 550 $\milli\joule$. The diameter of the nozzle is 2 $\milli\meter$, but the gas column is a little bit wider than this due to expansion. The plasma density measured by the interferometer shows that the density profile can be approximated by an approximately 400 $\micro\meter$ linear up ramp, followed by a 600 $\micro\meter$ plateau and a 1.2 $\milli\meter$ linear down ramp. The laser focal plane is placed 350 $\micro\meter$ upstream from the nozzle edge and 700 $\micro\meter$ above the nozzle exit. A 0.8 Tesla dipole consisting of two permanent magnets with a gap of 1.5 $\centi\meter$ is placed 60 $\centi\meter$ downstream from the nozzle and disperses the accelerated bunch in the plane of polarization of the laser pulse. A 2-mm-thick lead plate with a 2-mm-wide slit is placed in front of the dipole. A thin aluminum foil is placed in front of the slit to reflect the residual laser pulse, and another aluminum foil is placed in front of the lanex screen to shield it from the scattered light. The light emitted from the lanex screen is imaged onto a CCD camera (not shown in Fig. \ref{fig:4}).

In the experiment, the laser and plasma parameters were scanned without the dipole and slit to optimize the accelerator. At an optimal condition, the pointing stability of the electron bunch was measured to be 2.1 $\milli\radian$ along the laser polarization direction. The fluctuation of the bunch location caused by this pointing jitter is 1.15 mm, given that the distance between the nozzle and the slit is 0.55 m. As a consequence, the possibility for the bunch to go through the slit is 61\% since the width of the slit is 2 mm. The phosphor screen (Lanex) and the imaging system are well calibrated, thus the charge of the bunch can be calculated from the CCD count using a similar procedure described in Ref. \cite{Wu:2012ki}. The measured charge varies from 2 to 10 pC, depending on the laser and plasma parameters. Then the dipole and the slit were inserted to measure the energy spectrum. The energy resolution in the range from 30 to 90 $\mega\electronvolt$ is better than 1\%.

\begin{figure}[btp]
\includegraphics[width=0.4\textwidth]{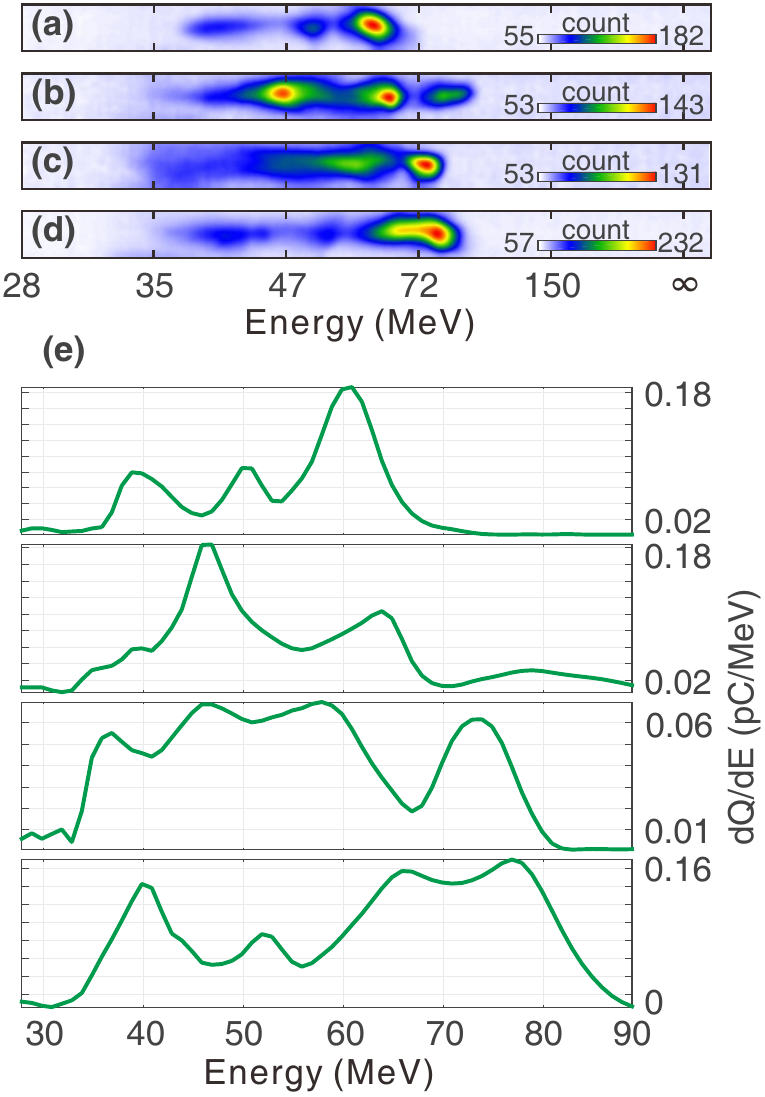}
\caption{\label{fig:5} (a)-(d) are the raw experimental spectra as seen on the Lanex screen. The color bars represent for the charge density. The corresponding linearized energy spectrums are shown in (e).}
\end{figure}

Four selected energy spectra are shown in Fig. \ref{fig:5}(a)-(d). These shots are taken under similar laser ($448\pm48~\milli\joule$) and plasma ($1.26\pm0.11\times10^{19}~\centi\meter^{-3}$) parameters. Fig. \ref{fig:5}(e) shows the corresponding (from top to bottom) deconvoluted spectra. It can be seen clearly in Fig. \ref{fig:5}(a)-(d) and the corresponding subfigures in Fig. \ref{fig:5}(e) that the energy spectrums have multiple peaks. Although it is possible for the spectrum to have a peaked structure without the laser-bunch interaction, the fact that the peaks in the energy spectrums are almost equally separated gives us confidence that those peaked structures are highly likely caused by the laser-bunch interaction and the subsequent energy filtering provided by the slit. These equally separated peaks also indicate that the energy chirp of these bunches are linear, as illustrated in Fig. \ref{fig:3}. In fact, we used equal energy spacing as a criterion to select those shots where the energy chirps are likely to be linear so that their current profiles can be reconstructed.

\begin{figure}[btp]
\includegraphics[width=0.45\textwidth]{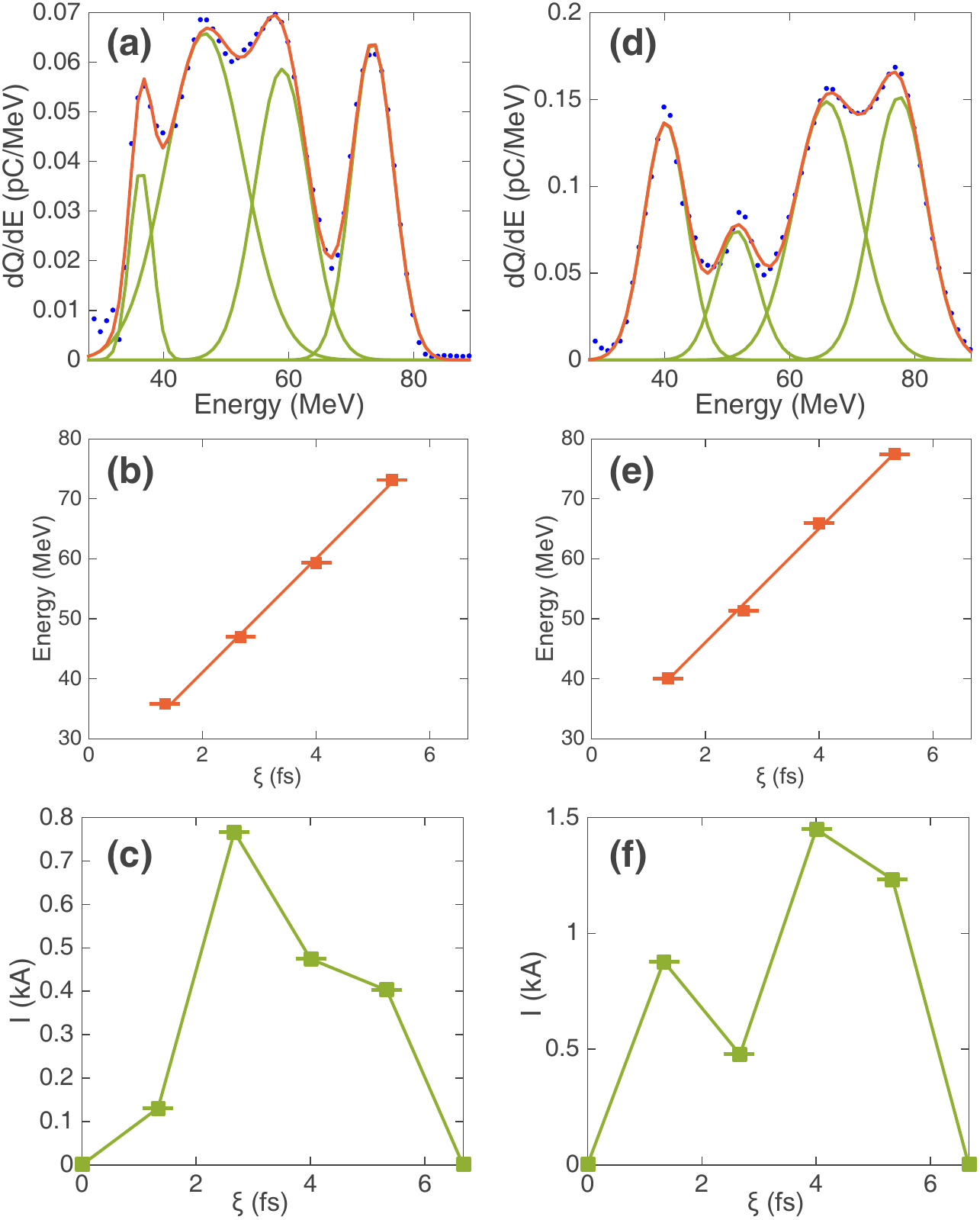}
\caption{\label{fig:6} (a) shows a typical deconvoluted energy spectrum (the blue dots) fitted by four Gaussian peaks (the solid green lines), where each peak represents a single beamlet. The red line is the superposition of the fitted peaks. (b) The recovered longitudinal phase space. (c) The recovered current profile. The deconvoluted energy spectrum, the recovered longitudinal phase space and the recovered current profile for another shot with higher density are shown in (d)-(f), respectively. The error bars in (b), (d), (e) and (f) represent for the uncertainty of bunchlet separation (see main text).}
\end{figure}

For these shots where the energy chirps are linear, we can assume that each peak in the deconvoluted energy spectrum represents a single bunchlet. In Fig. \ref{fig:6}(a) taken at a plasma density of $1.1\times10^{19}~\centi\meter^{-3}$, the energy spectrum contains four well separated peaks as shown by the blue dots. As before the four peaks are fitted by Gaussian profiles, as shown by the green solid lines. The central energy for each bunchlet can be found from the fitting data easily. The separation between neighboring peaks in the real space is half the laser wavelength. Thus the energy chirp can be mapped out as shown in Fig. \ref{fig:6}(b). The energy chirp is indeed linear as expected.

When recovering the energy chirp, the separation of neighboring bunchlet is assumed to be half the original laser wavelength. Actually the wavelength of the back part of the laser pulse may change as a result of laser-plasma interaction\cite{Esarey:2009ks,Geddes:2005kx,Murphy:2006gu}. However, simulations show that the laser frequency variation is typically less than 10\%. This effect was not measured in this experiment. Instead we have used the original wavelength (0.8 $\micro\meter$) as an approximation.

Since the energy chirp of the electron bunch is linear, the length of each bunchlet is proportional to the width of the corresponding peak in Fig. \ref{fig:6}(a). Given that the total charge of each bunchlet is equal to the area of the corresponding peak, the current for each bunchlet can be calculated. Therefore, the beam current profile can be recovered given that the separation of adjacent bunchlets is half the laser wavelength. Figure \ref{fig:6}(c) shows the calculated current profile which corresponds to the energy spectrum shown in Fig. \ref{fig:6}(a).

The horizontal error bars in Fig. \ref{fig:6}(b) and (c) arise from the uncertainty in determining the bunchlet separation. The pointing jitter of the electron bunches along the laser polarization direction is $\sigma_\theta=2.1~\milli\radian$, which leads to a fluctuation of the bunch location of $\sigma_x=1.15~\milli\meter$, given that the distance between the nozzle and the slit is 0.55 m. The fluctuation of the bunch location can affect the calculation of the separation of the adjacent bunchlets as explained before. The spot size of the bunch is $\sigma_b=3.5~\milli\meter$. Thus the change of the bunchlet separation caused by the fluctuation of the bunch location is $\frac{2}{\pi}\sin^{-1}(\sigma_x/\sigma_b)\approx0.1\lambda_0$, which means that the uncertainty of the separation between adjacent bunchlet is 0.28 fs, as shown by the horizontal error bars in Fig. \ref{fig:6}.

Figure \ref{fig:6}(d)-(f) show the energy spectrum, the recovered energy chirp and the recovered current profile for another shot with a higher plasma density of $1.31\times10^{19}~\centi\meter^{-3}$. The current profile in Fig. \ref{fig:6}(f) is more complex than that in Fig. \ref{fig:6}(c), illustrating the dependence of the beam current profile on the laser and plasma parameters. The peak currents in these two cases are 0.8 kA and 1.5 kA, respectively. The total charge for these bunches cannot be measured simultaneously due to the placement of the slit. However, we can use the total charge measured without the slit under the similar laser and plasma parameters as a reference. For the lower density case shown in Fig. \ref{fig:6}(a), the measured total charge is $4.0\pm1.7$ pC (average of 13 shots). While for the higher density case, the total charge is $5.3\pm2.1$ pC (average of 36 shots). The integration of the recovered current profiles in Fig. \ref{fig:6}(c) and (f) gives total charges of 2.4 and 5.4 pC respectively, which agree reasonably with the measured total charge.

It should be noted here that this method does not have the capability to distinguish whether the energy chirp is positive or negative. As a result, both the energy chirp and the current profile will still be consistent with the experiment observations even if they are reversed along the beam propagation direction. 

As a concept demonstration, we use only one laser pulse to drive the wake and to deflect the electron bunch at the same time. This is not the perfect configuration because it provides less flexibility and may decrease the transverse beam quality. By using a second laser to deflect the bunch, the flexibility and reliability of this method can be improved. Similar to the traditional deflecting cavity technique, the method described here is also destructive in a sense that it relies on the modification of the transverse phase space of the bunch by the laser field.

\section{\label{sec:conclusions}Conclusions}
A new method for mapping the current profile of electron bunches with linear chirps generated from a laser wakefield accelerator is described. Results of 3D PIC simulations and experiments are presented that validate this method. Ultrashort electron bunches with central energy of $\sim65$ MeV are generated via ionization-injection using a nominally 500 mJ, 40 fs laser combined with a helium-nitrogen gas mixture target. By analyzing the measured energy spectrum, the shots where the electron bunch is likely to have a linear energy chirp can be selected. As a consequence, the amplitude of the energy chirp and the current profile of these bunches are recovered. The peak current is on the order of 1 kA with an rms bunch length of 1.8 fs.

\begin{acknowledgments}
The authors thank Dr. Te-Sheng Hung, Mr. Ying-Li Chang and Mr. Yau-Hsin Hsieh for operating the laser. This work was supported by the National Basic Research Program of China Grant No. 2013CBA01501, NSFC grants 11425521, 11535006, 11175102, 11005063 and the Thousand Young Talents Program. The work at UCLA was funded by NSF grant PHY-145386 and DOE grant DE-SC0008491.
\end{acknowledgments}

\bibliography{measurement_of_current_profile}

\end{document}